\documentclass[english]{article}
\usepackage[utf8]{inputenc}
\usepackage[T1]{fontenc}
\usepackage{babel}
\usepackage{amsmath}
\usepackage{graphicx}
\usepackage{fancyhdr}
\usepackage{siunitx}
\usepackage{xcolor}
\usepackage{soul}
\usepackage{url}

\newcommand{\intinterval}{\mathrel{{.}\,{.}}\nobreak}

\begin{document}

\title{A dataset of continuous affect annotations and physiological signals for emotion analysis}

\author{Karan Sharma\textsuperscript{1{*}}, 
		Claudio Castellini\textsuperscript{1},
		Egon L. van den Broek\textsuperscript{2}, \\
		Alin Albu-Schaeffer\textsuperscript{1} \&
		Friedhelm Schwenker\textsuperscript{3}}

\maketitle
\thispagestyle{fancy}

\noindent1. Institute of Robotics and Mechatronics, DLR -- German Aerospace Center, Wessling, Germany. \\
2. Department of Information and Computing Sciences, Utrecht University, Utrecht, The Netherlands. \\
3. Institute of Neural Information Processing, Ulm University, Ulm, Germany. \\
{*}corresponding author(s): Karan Sharma (karan.sharma@dlr.de)

\begin{abstract}

From a computational viewpoint, emotions continue to be intriguingly hard to understand. In research, direct, real-time inspection in realistic settings is not possible. Discrete, indirect, post-hoc recordings are therefore the norm. As a result, proper emotion assessment remains a problematic issue. The Continuously Annotated Signals of Emotion (CASE) dataset provides a solution as it focusses on real-time continuous annotation of emotions, as experienced by the participants, while watching various videos. For this purpose, a novel, intuitive joystick-based annotation interface was developed, that allowed for simultaneous reporting of valence and arousal, that are instead often annotated independently. In parallel, eight high quality, synchronized physiological recordings (1000Hz, 16-bit ADC) were made of ECG, BVP, EMG (3x), GSR (or EDA), respiration and skin temperature. The dataset consists of the physiological and annotation data from 30 participants, 15 male and 15 female, who watched several validated video-stimuli. The validity of the emotion induction, as exemplified by the annotation and physiological data, is also presented.


\end{abstract}

%

\section*{Background \& Summary}
\label{sec:background}

The field of \emph{Artificial Intelligence} (AI) has rapidly advanced in the last decade and is on the cusp of transforming several aspects of our daily existence. For example, services like customer support and patient care, that were till recently only accessible through human--human interaction, can nowadays be offered through AI-enabled conversational chatbots~\cite{Oord2016} and robotic daily assistants~\cite{Hagengruber2017}, respectively. These advancements in interpreting explicit human intent, while highly commendable, often overlook implicit aspects of human--human interactions, especially the role emotions play in the same. Addressing this shortcoming is the aim of the interdisciplinary field of \emph{Affective Computing} (AC, also known as \emph{Emotional AI}), that focuses on developing machines capable of recognising, interpreting and adapting to human emotions~\cite{Picard1995, Mcstay2018}.

A major hurdle in developing these \emph{affective} machines is the internal nature of emotions that makes them inaccessible to external systems~\cite{Broek2010}. To overcome this limitation, the standard AC processing pipeline~\cite{Broek2011} involves: (\romannumeral 1) acquiring measurable indicators of human emotions, (\romannumeral 2) acquiring subjective annotations of internal emotions, and (\romannumeral 3) modelling the relation between these indicators and annotations to make predictions about the emotional state of the user. For undertaking steps (\romannumeral 1) and (\romannumeral 2) several different strategies are used. For example, during step (\romannumeral 1) different modalities like, physiological signals~\cite{Broek2010,Hanke2016,Gatti2018}, speech~\cite{Schwenker2017a} and facial-expressions~\cite{Taamneh2017, Soleymani2011} can be acquired. Similarly, the approaches to step (\romannumeral 2) usually vary along the following two main aspects. First, on the kind of annotation scale employed, i.e., either discrete or continuous. Second, on the basis of the emotion-model used, i.e., either discrete emotion categories (e.g., joy, anger, etc.) or dimensional models (e.g., the \emph{Circumplex model}~\cite{Russell2003}). Traditionally, approaches based on discrete emotional categories were commonly used. However, these approaches were insufficient for defining the strength~\cite{Gatti2018, Broek2011, Soleymani2011} and accounting for the time-varying nature~\cite{Soleymani2015} of emotional experiences. Therefore, nowadays continuous annotation based on dimensional models is preferred and several annotation tools for undertaking the same have been developed~\cite{Cowie2000, Nagel2007, Sharma2017}. Notwithstanding these efforts at improving the annotation process, a major impediment in the AC pipeline is that both, steps (\romannumeral 1) and (\romannumeral 2), require direct human involvement in form of subjects from whom these indicators and annotations are acquired~\cite{Gatti2018, Kaechele2016}. This makes undertaking these steps a fairly time-consuming and expensive exercise.

To address this issue, several (uni- and multi-modal) datasets that incorporate continuous annotation have been developed. Principal among these are the DEAP~\cite{DEAP}, SEMAINE~\cite{SEMAINE}, RECOLA~\cite{RECOLA}, DECAF~\cite{DECAF} and SEWA~\cite{AVEC2017}. The annotation strategy used in these datasets have the following common aspects. First, the two dimensions of the Circumplex model (i.e., \emph{valence} and \emph{arousal}) were annotated separately. Second, in all datasets except SEWA, that uses a joystick, mouse-based annotation tools were used. In recent years, both these aspects have been reported to have major drawbacks~\cite{Metallinou2013, Baveye2015, AVEC2017}. These being, that separate annotation of valence and arousal doesn't account for the inherent relationship between these dimensions~\cite{Metallinou2013, Girard2017}, and that mouse-based annotation tools are generally less ergonomic than joysticks~\cite{Metallinou2013, Baveye2015, Yannakakis2015, AVEC2017}. To address these drawbacks, we developed a novel \emph{Joystick-based Emotion Reporting Interface} (JERI) that facilitates simultaneous annotation of valence and arousal~\cite{Antony2014, Sharma2016, Sharma2017, Sharma2018}. A testament to the efficacy of JERI is that in recent years, several similar annotation setups have been presented~\cite{Girard2017, Mitsuhiko2017}. However, currently there are no openly available datasets that feature JERI-like setups.

To address this gap, we developed the presented \emph{Continuously Annotated Signals of Emotion} (CASE) dataset (Data Citation 1). It contains data from several physiological sensors and continuous annotations of emotion. This data was acquired from 30 subjects while they watched several video-stimuli and simultaneously reported their emotional experience using JERI. The physiological measures included in the dataset are from Electrocardiograph (ECG), Blood Volume Pulse (BVP), Galvanic Skin Response (GSR), Respiration (RSP), Skin Temperature (SKT) and Electromyography (EMG) sensors. The annotation data has been previously used for several publications aimed at introducing, analysing and validating this approach to emotion annotation~\cite{Sharma2016, Sharma2017, Sharma2018}. However, it has not been previously released. To the best of our knowledge, this is the first dataset that features continuous and simultaneous annotation of valence and arousal, and as such can be useful to the wider Psychology and AC communities.

\section*{Methods}

\subsection*{Participants}
Thirty volunteers (15 males, age 28.6$\pm$4.8 years and 15 females, age 25.7$\pm$3.1 years; range of age 22-37 years) from different cultural backgrounds participated in the data collection experiment. These participants were recruited from an organisation-wide call for volunteers sent at the Institute of Robotics and Mechatronics, DLR. Upon registering for the experiment, an email containing general information and instructions for the experiment was sent to the participants. In this email, they were asked to wear loose clothing and men were asked to preferably shave facial hair, to facilitate the placement of sensors. All participants had a working proficiency in English and were communicated to in the same. More information on the sex, age-group, etc., of the participants is available in the metadata to the dataset (Data Citation 1).  

\subsection*{Ethics Statement}
This experiment is compliant with the World Medical Association's Declaration of Helsinki, that pertains to the ethical principles for medical research involving human subjects, last version, as approved at the 59th WMA General Assembly, Seoul, October 2008. Data collection from participants was approved by the institutional board for protection of data privacy and by the work council of the German Aerospace Center. A physician is part of the council that approved the experiment.

\subsection*{Experiment Design}
The experiment was setup with a \emph{within subjects design}. Accordingly, \emph{repeated measures} were made and all participants watched and annotated the different video-stimuli used for the experiment. To avoid carry-over effects, the order of the videos in a viewing session was modified in a pseudo-random fashion, such that the resulting video sequence was different for every participant. To isolate the emotional response elicited by the different videos, they were interleaved by a two-minute long blue screen. This two-minute period also allowed the participants to rest in between annotating the videos. More information on the video-sequences is available in the dataset (Data Citation 1).

\subsection*{Experiment Protocol}
\begin{figure*}[!bt]
  	\centering
  	\includegraphics[width=\textwidth]{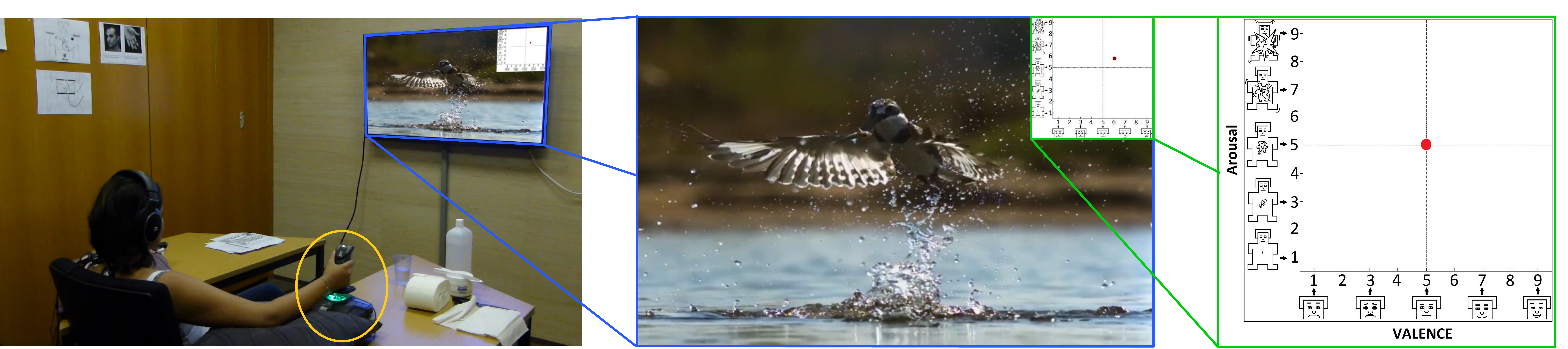}
  	\caption{The typical experiment setup (left) shows, a participant watching a video and annotating using JERI (joystick circled). The video-playback window (center) with the embedded annotation interface, that has the Self-Assessment Manikin (SAM) added to the valence and arousal axes (right).}
  	\label{fig:jeri}
\end{figure*}

On the day of the experiment, the participants were provided an oral and a written description of the experiment. After all questions regarding the experiment were addressed, the participants were asked to sign the informed consent form. Then, a brief introduction to the 2D circumplex model was provided and any doubts about the same were clarified. Following this, physiological sensors were attached and the participant was seated facing a 42'' flat-panel TV (see Figure \ref{fig:jeri}, left). Detailed information was then provided on the annotation procedure. It was emphasised to the participants that they should annotate their emotional experience resulting from the videos, and not the emotional content of the videos. To accustom the participants to the annotation interface, they were asked to watch and annotate five short practice videos. During this practice session, the experiment supervisor intervened whenever the participant asked for help and if required, provided suggestions at the end of every video. This session was also used to inspect the sensor measurements and if required, make appropriate corrections. After the practice session, the experiment was initiated and lasted for approximately 40 minutes. At the end of the experiment, feedback on the annotation system was acquired using the SUS questionnaire~\cite{Sharma2017,Sauro2011b}. Then, the sensors were removed and refreshments were offered. The participants were also encouraged to share any further insights they had on the experiment.

\subsection*{Annotation Interface}
\begin{figure*}[!bt]
  	\centering
  	\includegraphics[width=\textwidth]{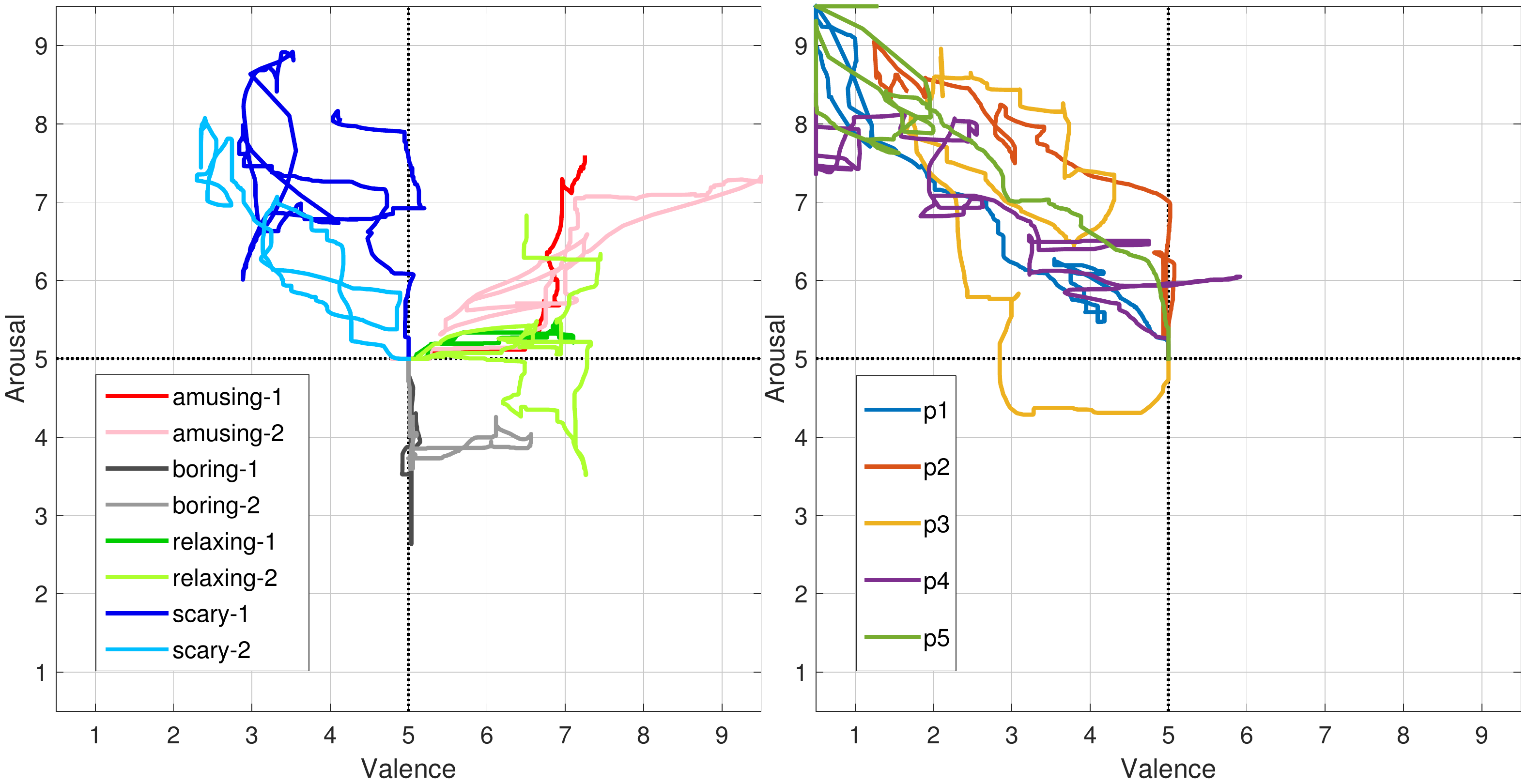}
  	\caption{The plot on the left shows the annotations from one participant for the different videos (see Table \ref{tab:videos}) in the experiment. The annotations for the `scary-2' video by the first five participants (labelled as p1--p5) can be seen in the plot on the right.}
  	\label{fig:annotations}
\end{figure*}

Figure \ref{fig:jeri} (right) shows the design of the annotation interface. It is based on the 2D circumplex model that has been supplemented with the Self-Assessment-Manikin (SAM)~\cite{Bradley1994} on its co-ordinate axes. These manikin depict different valence (on X--axis) and arousal (on Y--axis) levels, thereby serving as a non-verbal guide to the participants during annotation. The red pointer in the figure shows the resting/neutral position. The participants were instructed to annotate their emotional experience by moving/holding the red pointer in the appropriate region of the interface. Since the annotation was done over the entire length of a video, it results in a continuous 2D trace of the participant's emotional experience (see Figure \ref{fig:annotations}). The position of the annotation interface inside the video-playback window is shown in Figure \ref{fig:jeri} (center). This position can be easily changed, but since none of the participants requested that, it was retained as shown for all participants. The annotation interface was developed in the National Instruments (NI) LabVIEW programming environment.

\subsection*{Videos}
In this experiment, the aim was to elicit multiple emotional states, namely \textit{amusing, boring, relaxing} and \textit{scary}, through video-stimuli. To this end, 20 videos previously used by other studies were shortlisted~\cite{Gross1995, Hewig2005, Bartolini2011}. The emotional content of these videos was then verified in a pre-study, where 12 participants (no overlap with the participants of this study) viewed and rated these videos remotely using a web-based interface. Based on the results of this pre-study and further internal reviews, eight videos were selected for the main experiment, such that there were two videos each for every emotional state that we wanted to elicit. Additionally, three other videos were also used in the experiment, i.e., the start-video, the end-video and the interleaving blue-screen videos. More information on the videos is available in Table \ref{tab:videos}, in the Usage Notes section and in the dataset (Data Citation 1).

\begin{table}[!ht]
\centering
\resizebox{\textwidth}{!}{
\begin{tabular}  {| l | l | c | c | c | c |}
\hline
    Source						& Video-Label & Video-ID & \multicolumn{2}{|c|}{Intended Attributes} & Dur. $[s]$ \\ \cline{4-5} 
	            				&             &          & Valence      & Arousal                    &         \\ \hline \hline
	Hangover				   	& amusing-1   & 1        & med/high     & med/high                   & $ 185 $ \\ \hline
	When Harry Met Sally		& amusing-2   & 2        & med/high     & med/high                   & $ 173 $ \\ \hline	
	European Travel Skills	 	& boring-1 	  & 3		 & low          & low                        & $ 119 $ \\ \hline
	Matcha: The way of Tea 		& boring-2    & 4        & low          & low                        & $ 160 $ \\ \hline
	Relaxing Music with Beach	& relaxing-1  & 5        & med/high     & low                        & $ 145 $ \\ \hline
	Natural World: Zambezi		& relaxing-2  & 6        & med/high     & low                        & $ 147 $ \\ \hline
	Shutter						& scary-1	  & 7 		 & low          & high                       & $ 197 $ \\ \hline
	Mama						& scary-2     & 8        & low          & high                       & $ 144 $ \\ \hline
	Great Barrier Reef			& startVid	  & 10       & - 			& -			              	 & $ 101 $ \\ \hline
	Blue screen with end credits& endVid      & 12       & - 			& -			               	 & $ 120 $ \\ \hline
	Blue screen					& bluVid      & 11       & - 			& -			                 & $ 120 $ \\ \hline
\end{tabular}}
\caption{The source, label, ID used, intended valence-arousal attributes and the duration of the videos used for the dataset.}
\label{tab:videos}
\end{table}

\subsection*{Sensors \& Instruments}
The physiological sensors used for the experiment were selected based on their prevalence in AC datasets and applications~\cite{Picard1995,Broek2011,DEAP,Taamneh2017}. Other sensors and instruments were chosen based on either the recommendations of the sensor manufacturer or on how interfaceable they were with the data acquisition setup. More details on these sensors and instruments are provided in this subsection and Table \ref{tab:sensors}. 

\textbf{ECG sensor.} The electrical signal generated by the heart muscles during contraction can be detected using an ECG sensor. The procedure  used involves placement of three electrodes in a triangular configuration on chest of the participant. Two electrodes are placed on the right and left \emph{coracoid processes} and the third on the \emph{xiphoid process}~\cite{TTtut}. This sensor also pre-amplifies and filters the detected electric signal.

\textbf{BVP sensor.} Also known as a Photoplethysmography (PPG) sensor, it emits light into the tissue and measures the reflected light. The amount of observed reflected light varies according to the blood flowing through the vessels, thus serving as a measure for cardiac activity. This sensor was placed on the middle finger of the non-dominant hand~\cite{TTtut}.

\textbf{GSR sensor.} Also known as Electrodermal Activity (EDA) sensor, it measures the variation in electrical conductance resulting from sweat released by the glands on the skin. The two electrodes emanating from this sensor were placed on the index and ring fingers of the non-dominant hand~\cite{TTtut}. 

\textbf{Respiration sensor.} The expansion and contraction of the chest cavity can be measured using a Hall effect sensor placed around the \emph{pectoralis major} muscle~\cite{TTtut}. Thus, this sensor measures the respiration rate of the participant.

\textbf{Skin temperature sensor.} Small variations in skin temperature were measured and converted to electrical signals using an epoxy rod thermistor. This sensor was placed on the pinky finger of the non-dominant hand~\cite{TTtut}. 

\textbf{EMG sensors.} The surface voltage associated with muscle contractions can be measured using a surface-Electromyography (sEMG, simply EMG) sensor. Previous research in AC has generally focused on three muscles. These are the \emph{corrugator supercilii} and \emph{zygomaticus major} muscle groups on the face, and the \emph{trapezius} muscle on the upper-back. Accordingly, a total of three EMG sensors (one each for the aforementioned muscles) were used for the experiment. These sensors also pre-amplify and perform an analog Root Mean Square (RMS) conversion on the measured raw EMG signal~\cite{TTtut}. 

\textbf{Sensor isolators.} The sensor manufacturer recommends using a `sensor isolator' to ensure electrical isolation between the participants and the powered sensors. Accordingly, the physiological sensors were indirectly connected to the data acquisition module, through these sensor isolators (see Figure \ref{fig:setupschema}).

\textbf{Data acquisition modules.} A 32-channel (16-channel differential) Analog-to-Digital Conversion (ADC) module with 16-bit resolution was used to acquire the output voltages from the sensor isolators (indirectly, the sensors). This module is connected to a Data Acquisition (DAQ) system that transfers the data to the acquisition PC.        
 
\textbf{Joystick.} The joystick is the only instrument in the experiment that is directly controlled by the participants. The used joystick is a generic digital gaming peripheral that features a return spring. This provides the user proprioceptive feedback about the location of the pointer in the interface, thereby helping to mitigate the cognitive load associated with simultaneous tasks of watching the video and annotating her emotional experience~\cite{Sharma2017,Girard2017}.

\begin{table}[!ht]
\centering
\resizebox{\textwidth}{!}{
\begin{tabular}  {| l | l | l | l | l | c |}
\hline
    Sensor/Instrument	& No.	& Manufacturer			& Model 	& \multicolumn{2}{c|}{Conversions} \\ \cline{5-6}  
						&		&						&			& Equations									& Units \\ \hline \hline
	ECG sensor         	& 1  	& Thought Technology    & SA9306    & $V_{out} = (V_{in} - 2.8) / 50 \cdot 10^{3}$& $mV$ \\ \hline 
	BVP	sensor			& 1  	& Thought Technology    & SA9308M   & ${BVP} \% = 58.962V_{in} - 115.09$	& $\%$ 	\\ \hline
	GSR	sensor			& 1  	& Thought Technology    & SA9309M   & $G = 24V_{in} - 49.2$					& ${\mu}S$  \\ \hline	
	Respiration sensor	& 1 	& Thought Technology	& SA9311M   & $R\% = 58.923V_{in} - 115.01$			& $\%$	\\ \hline
	Skin temp. sensor	& 1    	& Thought Technology    & SA9310M   & $T = 21.341V_{in} - 32.085$			& $^{\circ}C$ \\ \hline
	EMG sensor			& 3  	& Thought Technology    & SA9401M-50& $V_{out}(RMS)=(V_{in}- 2) /4000 \cdot 10^{6}$ & ${\mu}V$ \\ \hline
	Sensor Isolator		& 2  	& Thought Technology    & SE9405AM  & -											& - \\ \hline
	ADC	module			& 1	  	& National Instruments  & NI 9205   & -											& - \\ \hline
	DAQ system			& 1     & National Instruments  & cDAQ-9181	& -											& - \\ \hline
	Joytick				& 1	  	& Thrustmaster       	& T.16000M	& $pos_{out} = 0.5+9\cdot(pos_{in}+26225)/52450$& - \\ \hline
\end{tabular}}
\caption{The type, number (No.), manufacturer and model of different sensors and instruments used in the experiment. Wherever applicable, the conversion equations used to transform the logged input values to the desired output units/scales (see last column) are also presented.}
\label{tab:sensors}
\end{table}

\subsection*{Data Acquisition}
\begin{figure*}[!bt]
  	\centering
  	\includegraphics[width=\textwidth]{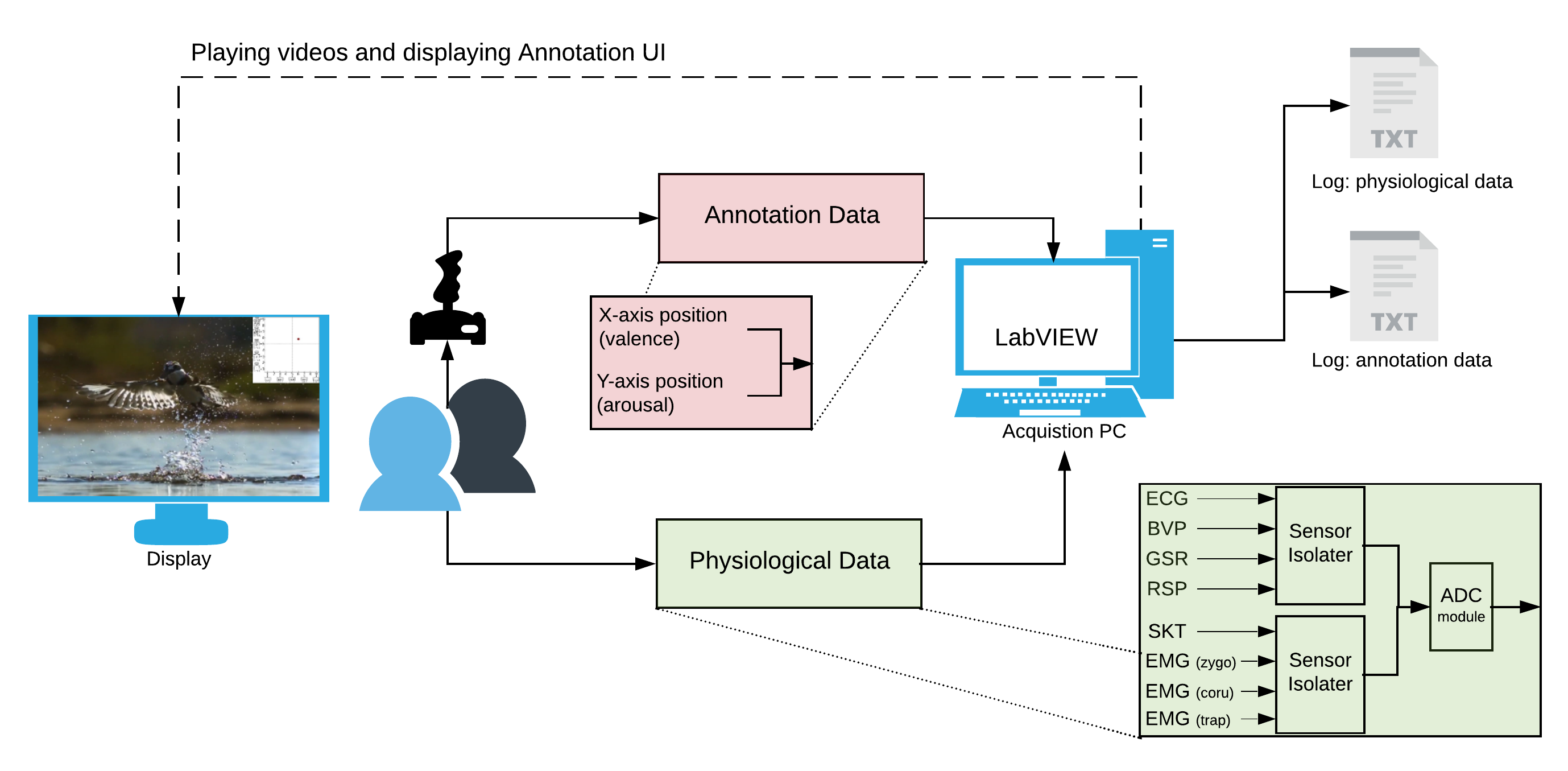}
  	\caption{The schematic shows the various aspects of the experiment and the data acquisition setup. The arrows indicate the direction of the data-flow. The solid and the dotted lines indicate the primary and secondary tasks of the acquisition process, respectively.}
  	\label{fig:setupschema}
\end{figure*}

Figure \ref{fig:setupschema} shows the experiment and the data acquisition setup. The video-playback, the annotation interface and the data acquisition components were all directly managed through LabVIEW. This allows for a seamless integration of all these different components. The open-source VLC media player was used for video-playback. The joystick was directly connected to the acquisition PC over a USB port. The physiological data was acquired over Ethernet using the DAQ system. The acquisition rate for the annotation and the physiological data was 20 Hz and 1000 Hz, respectively. The acquired data was augmented with the timestamp provided by the acquisition PC and logged in two different text files, i.e., one each for the physiological and the annotation data. The same process was repeated for all participants resulting in 60 ($30 \times 2$) log files.   

\subsection*{Data Preprocessing}
The procedure used for transforming the raw log files into the presented dataset is summarized in this subsection. In the following, step 1 was performed once and steps 2--5 were iteratively applied to log files for each participant.

\begin{enumerate}
\item \textit{Duration of the videos:} using the \textit{ffprobe} tool from the \emph{FFmpeg} multimedia framework~\cite{FFmpeg}, the exact duration (in milliseconds) of the videos was determined and has been made available in the dataset (Data Citation 1).

\item \textit{Transforming raw data:} the sensor input received by the sensor isolators gets modified before being transferred to the DAQ system. To rectify the effects of this modification, the logged voltages need to be transformed. This was achieved by applying the equations presented in Table \ref{tab:sensors} to yield the desired output with specific units/scales (Table \ref{tab:sensors}, last column). Similarly, the logged annotation data, which was in the integer interval $[-26225 \intinterval 26225]$, was also rescaled to the annotation interface interval $[0.5 \intinterval 9.5]$ by using the equations presented in Table \ref{tab:sensors}. 

\item \textit{Data interpolation:} a common problem in data acquisition and logging is the latencies that can be introduced during any of these processes. This was also evident in our data, where, e.g., the time between the subsequent samples of the annotation data was occasionally more than the expected 50 ms. To address this issue, linear interpolation was performed on the physiological and the annotation data. For undertaking the same, first, two time-vectors with sampling intervals of 1 ms (for the physiological data) and 50 ms (for the annotation data) were generated based on the time-duration of the logged data. These vectors serve as the query points for the interpolater that determines the value at these points by fitting a line between the corresponding discrete samples in the logged data. As a result of the interpolation process, the resulting sampling intervals for the physiological and the annotation data were 1 ms and 50 ms, respectively. For the eventuality that other researchers might prefer to use either the non-interpolated data or different interpolation methods, the original non-interpolated data is also available in the dataset (Data Citation 1).     

\item \textit{Adding the video-IDs:} the log files contain timestamps, but do not have information identifying the duration and the order of the videos. Hence, the extracted video-durations and a lookup table of the video-sequences were used to identify the data segments pertaining to each video. Then, this information was added as an extra column to the log files, containing the different video-IDs (see Table \ref{tab:videos}). This process was also undertaken for the non-interpolated data.

\item \textit{Saving the data:} the resulting data from the aforementioned steps was saved into two different \emph{comma-separated values} (csv) files, i.e., one each for the physiological and the annotation data. The csv format was chosen as it is natively accessible by different programming and scientific computing frameworks. 
\end{enumerate}

\subsection*{Code availability}
The LabVIEW-based graphical code for the experiment and data acquisition is highly specific to the sensors and equipment used in our experiment. It has therefore not been made available with this dataset (Data Citation 1). Nevertheless, readers who wish to replicate the experiment can contact the corresponding author for further assistance. For readers who want to reproduce the experiment, we hope the detailed description provided in this article will suffice.  

The data preprocessing steps outlined in the previous subsection were implemented in MATLAB 2014b. The linear interpolation was performed using the \emph{interp1} function. The raw log files, that the data preprocessing code acts upon, are not a part of the released dataset because they do not contain the video-IDs. Hence, the preprocessing code has also not been released. Nevertheless, both the raw data and the preprocessing code, are available to interested researchers upon request.

%
\section*{Data Records}

The presented CASE dataset comprises of the processed data resulting from the aforementioned experiment. This dataset is hosted as a single archive file on the \emph{figshare} data repository (Data Citation 1) and is organised into three main directories. These are, (\romannumeral 1) the \textit{interpolated}, (\romannumeral 2) the \textit{non-interpolated} and (\romannumeral 3) the \textit{metadata} directories. At the root and the subsequent sub-directories, detailed \textit{README} files explaining the contents of these directories have been provided.          

\subsection*{metadata}
This directory contains auxiliary information about the experiment organised into the following three files:
\begin{enumerate}
\item \textit{participants.xlsx:} this Excel file contains the participant-ID, the sex, the age-group and the ID of the video-sequence used, for all participants in the experiment. 

\item \textit{video\_sequences.xlsx:} the video-stimuli were shown in a unique sequence to every participant. The columns of this Excel file contain video-IDs that indicate the ordering of the video-stimuli in these sequences.   

\item \textit{videos.xlsx:} in addition to the attributes already presented in Table \ref{tab:videos}, this Excel file contains further information on the used video-stimuli. This includes, the videos' durations in milliseconds, links to the IMDb/YouTube entries for the videos' sources, URLs to the videos and the time-window for the videos at these URLs. More information on how to acquire these videos is presented in the Usage Notes section. 
\end{enumerate}

\subsection*{interpolated and non-interpolated}
Both of these directories share similar structure and filenaming conventions, with the only difference being the process used to generate the files contained in them (see the Subsection on Data Processing). Hence, unless stated otherwise, the description provided here is applicable to the files in both of these directories. The data contained in these directories is organised into two sub-directories. These being, (\romannumeral 1) the \textit{annotations} and (\romannumeral 2) the \textit{physiological} directories containing the participant-wise annotation and physiological data, respectively. An overview of the data records in these sub-directories is provided below.  

\subsubsection*{annotations/sub\_XX.csv}
This directory contains 30 annotation files titled \texttt{sub\_XX.csv}, where \texttt{XX} are natural numbers in the set $\{1,2,\dots,30\}$ denoting the IDs of the participants. The column-name and the content for the four columns in each csv file are as follows:
\begin{itemize}
\item \textit{Column 1: jstime.} Time in milliseconds from the beginning of the video-viewing session to the end. 
\item \textit{Column 2: valence.} The scaled X-axis value of the joystick position in the interface (see Table \ref{tab:sensors}).
\item \textit{Column 3: arousal.} The scaled Y-axis value of the joystick position in the interface (see Table \ref{tab:sensors}).
\item \textit{Column 4: video.} Contains the sequence of video-IDs that indicates the ordering and duration of the different video-stimuli for the given participant.  
\end{itemize}

\subsubsection*{physiological/sub\_XX.csv}
This directory contains 30 physiological data files titled \texttt{sub\_XX.csv}, where \texttt{XX} are natural numbers in the set $\{1,2,\dots,30\}$ denoting the IDs of the participants. The column-name and the content for the 10 columns in each csv file are as follows:
\begin{itemize}
\item \textit{Column 1: daqtime.} Time in milliseconds from the beginning of the video-viewing session to the end. 
\item \textit{Columns 2--9: ecg, bvp, gsr, rsp, skt, emg\_zygo, emg\_coru and emg\_trap.} The transformed sensor output values for each of 8 physiological sensors used in the experiment. More information on the sensors, the transformations applied, and the outputs units for these values, is available in Table \ref{tab:sensors} and the README files in these directories. 
\item \textit{Column 10: video.} Contains the sequence of video-IDs that indicates the ordering and duration of the different video-stimuli for the given participant.
\end{itemize}

\noindent \textbf{Note.} The \textit{jstime} and the \textit{daqtime} columns in the above mentioned files contain timestamps provided by a common clock on the logging computer. They have been named differently due to the different sampling intervals used for logging these files, i.e., 50 ms and 1 ms, respectively.

\section*{Technical Validation}

\subsection*{Annotation Data}
The quality and the reliability of the annotation data has been thoroughly addressed in our previous works~\cite{Antony2014,Sharma2016,Sharma2017,Sharma2018}. A summary of the relevant highlights from these works is presented below. 

In \cite{Antony2014,Sharma2016} several different exploratory data analyses were presented. These analyses provided an initial intuition into the annotation patterns for the different video-stimuli. For example, the annotations for the two scary videos had in general low valence and high arousal values. They were thus different from annotations for the amusing videos which had relatively high valence and medium arousal. These differences can also be seen in the annotations presented in Figure \ref{fig:annotations} (left). The initial exploratory results presented in \cite{Antony2014,Sharma2016} were then formally validated in \cite{Sharma2017}, where Multivariate ANOVA (MANOVA) was used to quantify the statistical significance of the differences in the annotations for these videos. The `usability' of our annotation approach was validated using the \emph{System Usability Scale} (SUS) questionnaire. According to the ratings received on the same, the annotation setup had `excellent' usability as the participants found it to be consistent, intuitive and simple to use~\cite{Sharma2017}. In \cite{Sharma2017,Sharma2018}, several different methods for analysing the annotation patterns in continuous 2D annotations were presented. The results of these continuous methods were concurrent to the results of the MANOVA. Also in \cite{Sharma2017,Sharma2018} several different methods for extracting additional information from these continuous annotations have been presented. For example, the \emph{Change Point Analysis} method in \cite{Sharma2017} automatically detects the major change-points in the annotation data that can be used to segment the annotations into several salient segments. For comparison with the physiological data, some results for the annotation data are presented in the next subsection.          

\begin{table}[!htb]
\centering
\resizebox{\textwidth}{!}{
\begin{tabular}  {| l | l |}
\hline
    Sensor				& Extracted Features   \\ \hline \hline  
	ECG          		& Heart Rate (HR)\\ & Inter-Beat Interval (IBI)\\ & Standard Deviation (SD) of NN-intervals (SDNN)\\ \hline 
	BVP					& Heart Rate (HR)\\ & Inter-Beat Interval (IBI)\\ & Standard Deviation (SD) of NN-intervals (SDNN)\\ \hline
	GSR					& Skin Conductance Level (SCL)\\ & Skin Conductance Response (SCR)\\ \hline	
	Respiration			& Respiration Rate (RR)\\ & Interval of Respiration peaks\\ \hline
	Skin Temperature 	& Temperature \\ & SD of Temperature (SDT)\\ \hline
	EMG--zygomaticus	& Amplitude of the signal \\ \hline
	EMG--corrugator		& Amplitude of the signal\\ \hline
	EMG--trapezius		& Amplitude of the signal\\ \hline
\end{tabular}}
\caption{The sensors and the various features extracted from the sensor signals.}
\label{tab:allfeats}
\end{table}

\subsection*{Physiological Data}
In the Background \& Summary section, the typical AC processing pipeline was presented. The final objective of this pipeline is to develop machine learning models that can infer the emotional state of humans from (in the given case) physiological signals. To achieve the same, it is critical that the physiological responses to the different video-stimuli are discernible from each other and are ideally correlated to annotation data. If indeed these patterns exist, they would validate the quality and the value of this data. To determine the same, we extracted several features from the physiological data and performed \emph{Principal Components Analysis} (PCA) on these features. The details and results of this analysis is presented as follows.

\subsubsection*{Feature Extraction}
The feature extraction was performed iteratively over the physiological data files for each participant. First, the data for a given participant was segmented into chunks for the different video-stimuli. Then, from the sensor data pertaining to each of these video-chunks, several features were extracted (see Table \ref{tab:allfeats}).

\begin{table}
\centering
\resizebox{\textwidth}{!}{
\begin{tabular}  {| l | l |}
\hline
    Sensor				& Feature Selected   \\ \hline \hline  
	ECG          		& mean HR\\ \hline 
	BVP					& Standard Deviation (SD) of NN-intervals (SDNN)\\ \hline
	GSR					& mean SCR\\ \hline	
	Respiration			& mean RR\\ \hline
	Skin Temperature 	& SD of Temperature (SDT)\\ \hline
	EMG--zygomaticus	& mean amplitude (mean Zygo)\\ \hline
	EMG--corrugator		& mean amplitude (mean Corr)\\ \hline
	EMG--trapezius		& mean amplitude (mean Trap)\\ \hline
\end{tabular}}
\caption{The sensors and the features selected from each sensor.}
\label{tab:selfeats}
\end{table}

\begin{figure}[!p] 
  	\centering
  	\includegraphics[width=\textwidth]{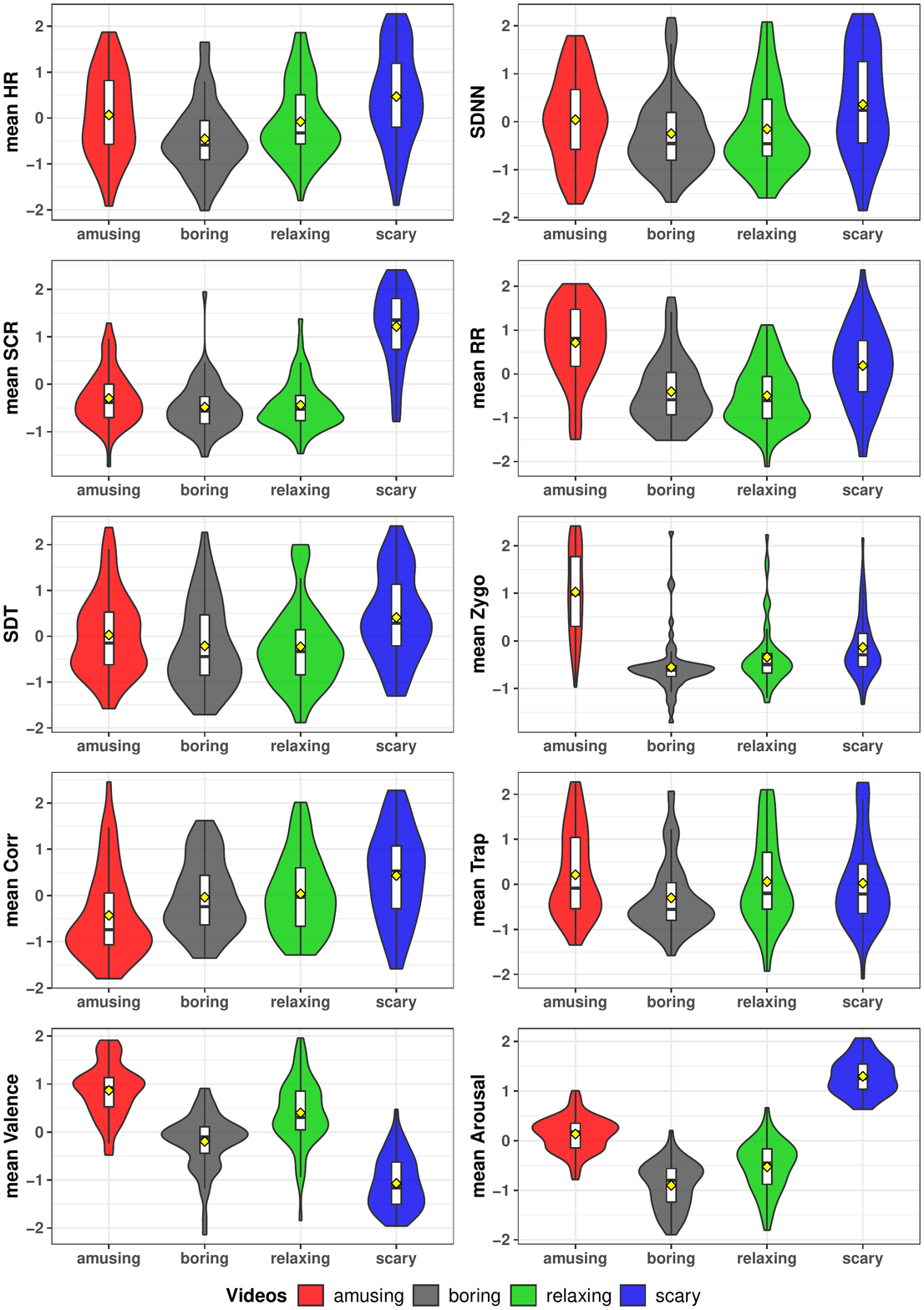}
  	\caption{"Violin" plots of the distribution of the selected features and the mean annotation (valence \& arousal) values across different types of videos. The box plots embedded in each violin plot show the Interquartile Range (IQR) for each considered variable, while a yellow diamond marks the mean of the distribution.}
  	\label{fig:violinplots}
\end{figure}

For the technical validation presented here, one predominantly used feature for each sensor was selected and where applicable, the mean of this feature across the given video-chunk was calculated. Similarly, the mean valence and arousal values across each video-chunk were calculated.  The selected physiological features are presented in Table \ref{tab:selfeats}. As a result, for the 30 participants who each watched eight emotional video-stimuli, we have 240 ($30 \times 8$) values for each of these selected features. Due to inter-personal differences, the participants have a different baseline value for each of these extracted features. These differences can be detrimental to the comparison of these features across all participants and were therefore removed using \emph{Z-score} standardisation across each participant. The same was also done for the annotation data. The violin-plots in Figure \ref{fig:violinplots} show the distributions of the selected features and annotation data, across the 4 different video-labels (see Table \ref{tab:videos}). From the figure, it is apparent that some of the physiological features (consider the top eight panels) characterise specific types of videos. For instance, scary videos result in high values of SCR and elevated HR, while amusing videos elicit accelerated respiration rates and activity of the \emph{zygomaticus} muscles. Boring and relaxing videos, as expected, elicit similar values of all features. These results are in line with previous research~\cite{Broek2010,Broek2011}, where, e.g., HR and SCR were determined to be positively correlated to arousal. This effect can also been seen in our data, where the reported arousal levels (see bottom-left panel) for scary videos are higher than for the other videos. Similarly, \emph{zygomaticus} activity which has been reported to be positively correlated to valence (see bottom-left panel), also exhibits similar patterns in our data.

\begin{figure}[!hbt]
  	\centering
  	\includegraphics[width=\textwidth]{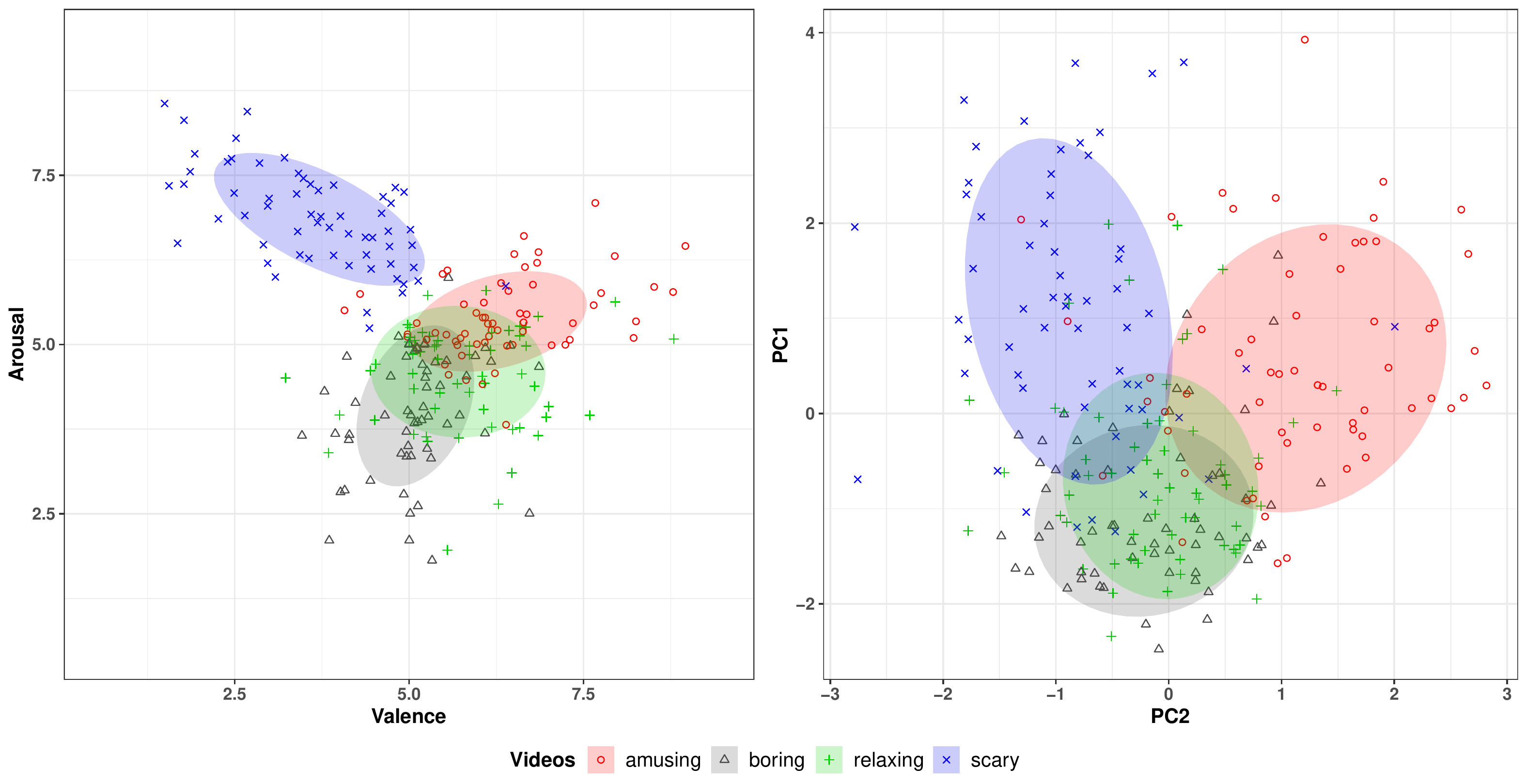} 
  	\caption{Scatter plots of the mean annotation data and the first two principal components of the physiological data, labelled according to the types of videos. Ellipses denote one standard deviation.}
  	\label{fig:pca}
\end{figure}

\subsubsection*{PCA} 
PCA is a commonly used dimensionality reduction technique~\cite{Jolliffe2005,Ringner2008}. This allows for the visualization of the given data in a lower dimensional space, where e.g., spatial distributions of the data can be analysed. The result of PCA for the selected \emph{Z-scored} features is shown in Figure \ref{fig:pca} (right), where the scores on the first two principal components are shown. The scatter plot on the left in Figure \ref{fig:pca} shows the mean valence and arousal values across the different video-labels. The data ellipses show the standard deviation of data pertaining to these video-labels. As is evident from this figure, the physiological and the annotation data form concurrent clusters.
These two figures validate the data in an even more prominent way than Figure \ref{fig:violinplots}. Valence and arousal values (left panel) of scary videos are concentrated in the upper-left quadrant, those for the amusing videos are in the upper-right, and the others lie in the middle with low arousal values, as one would expect. This is confirmed by the right panel, in which the four types of videos are represented analogously on the plane obtained using the first two principal components of the physiological features. This seems to indicate that the physiological features somehow "match" the joystick annotations. Of course, this serves only an initial investigation and a more rigorous analysis is required to fully exploit the potential of the database. Nevertheless, the results provided here show that the presented dataset has several viable characteristics that would make it of interest to our research community.

\section*{Usage Notes}

%

\subsection*{Videos}
Due to copyright issues, we cannot directly share the videos as a part of this dataset (Data Citation 1). Nonetheless, to help users in ascertaining the emotional content of these videos in more detail and if required, to replicate the experiment, we have provided links to websites where these videos are currently hosted (see /metadata/videos.xlsx). We are aware that these links might become unusable in the future and that this can cause inconvenience to the users. In such an eventuality, we encourage the users to contact us, so that we can assist them in acquiring/editing the videos.    

\subsection*{Feature Extraction and Downstream Analysis}
The code used for the technical validation of the dataset was developed in MATLAB 2014b and R-language (version 3.3.3). The feature extraction was done in MATLAB using open-source toolboxes/code like TEAP~\cite{TEAP} and an implementation of the Pan Tompkins QRS detector~\cite{QRSalgo}. The PCA analysis was performed in R using the \emph{prcomp} function from the \emph{stats} package. This code is available to interested researchers upon request. Users of the dataset (Data Citation 1) interested in leveraging the continuous nature of the provided annotations are advised to check our previous works~\cite{Sharma2017, Sharma2018}. The analysis presented in these works was primarily undertaken in R-language and can be easily reproduced. 
\section*{Data Citations}
1. Sharma, K., Castellini, C., van den Broek, E., Albu-Schaeffer, A. \& Schwenker, F. \textbf{CASE Dataset} (2018). Available at: \url{https://rmc.dlr.de/download/CASE_dataset/CASE_dataset.zip}

\section*{Acknowledgements}
The authors would like to thank the participants of the pre-study and the main experiment. We also want to thank Mr.\ Jossin Antony for co-developing the data acquisition software and assisting in data collection, and Dr.\ Freek Stulp for his help and support in developing this dataset.   

\subsection*{Author Contributions}

\hspace{\parindent}KS co-developed the data acquisition software, collected the data, developed the dataset, contributed to the technical validation and composed the manuscript.

CC co-designed the experiment, helped with participant enrolment, supervised the data collection, contributed to the technical validation and the composition of the manuscript.
 
EL assisted in the design and the development of the annotation interface, supervised the data processing and edited the manuscript.

AA verified the dataset and edited the manuscript.

FS co-designed the experiment, supervised the technical validation and edited the manuscript.

\section*{Competing financial interests}
The author(s) declare no competing financial interests.



\end{document}